**Anomalous Landau quantization in intrinsic magnetic topological insulators**

Su Kong Chong[1†*], Chao Lei[2†], Seng Huat Lee[3,4], Jan Jaroszynski[5], Zhiqiang Mao[3,4], Allan H. MacDonald[2], Kang L. Wang[1]*

[1]Department of Electrical and Computer Engineering, University of California, Los Angeles, California 90095, United States

[2]Department of Physics, The University of Texas at Austin, Austin, TX 78712

[3]2D Crystal Consortium, Materials Research Institute, The Pennsylvania State University, University Park, PA 16802, USA

[4]Department of Physics, The Pennsylvania State University, University Park, PA 16802, USA

[5]National High Magnetic Field Laboratory, Florida State University, Tallahassee, FL, USA

*Corresponding authors: sukongc@g.ucla.edu; wang@seas.ucla.edu

†These authors contributed equally to this work

**Abstract**

The intrinsic magnetic topological insulators, $Mn(Bi_{1-x}Sb_x)_2Te_4$, in their spin-aligned strong field configuration have been identified as a Weyl semimetal with single pair of Weyl nodes[1-4]. A direct consequence of the Weyl state is the layer-dependent Chern number (C) in thin film quantization. Previous reports in $MnBi_2Te_4$ thin films revealed the higher C states in the spin alignment by either increasing the film thickness[5] or controlling chemical potential into electron doping[6-8]. A clear picture of the higher Chern states is still missing as the situation is complicated by the emerging of surface band Landau levels (LLs) in magnetic field. Here, we report a tunable layer-dependent of C= 1 state with the Sb substitutions by performing a detailed analysis of the quantization states in $Mn(Bi_{1-x}Sb_x)_2Te_4$ dualgated devices, consistent with the calculations of the bulk Weyl point **separations in the compounds. The observed Hall quantization plateaus for our thicker Mn(Bi$_{1-x}$Sb$_x$)$_2$Te$_4$ films under strong magnetic fields can be interpreted from a theory of surface and bulk spin-polarized Landau levels spectrum in thin film magnetic topological insulators. Our results demonstrate that Mn(Bi$_{1-x}$Sb$_x$)$_2$Te$_4$ thin films provide a highly tunable platform for probing the physics of the anomalous Landau quantization that is strongly sensitive to magnetic order.**

**Introduction**

Magnetic topological insulators (MTIs) and Weyl semimetals have both received a great deal of attention in recent condensed matter physics research[9-11]. Particularly, the intrinsic MTI MnBi$_2$Te$_4$ has provided researchers with an ideal candidate to study the relationship between topological quantum states and magnetic phases[5-8,12-19]. Theoretical predictions[1-4] and recent experimental results[20] show that when its Mn local moment spins are aligned by an external magnetic field, the energy bands of bulk MnBi$_2$Te$_4$ have a single isolated pair of Weyl crossing points that are close to the Fermi level and therefore can be accessed by controlling the carrier-density.

In this work, we report on thickness-dependent magneto-transport studies of the mechanically-exfoliated Mn(Bi$_{1-x}$Sb$_x$)$_2$Te$_4$ thin flakes with three different Sb concentrations. The Sb substitutions in the MnBi$_2$Te$_4$ parent compound (*i*) move the Fermi level of the bulk bands closer to charge neutrality point (CNP)[20-22], and (*ii*) modulate the Weyl point separation in momentum space as illustrated in Fig.1a. We focus here on the Chern insulator states of the spin-moment aligned Mn(Bi$_{1-x}$Sb$_x$)$_2$Te$_4$. We show that Sb substitution extends the surface gap regime to a wider thickness range by suppressing conduction from the trivial bulk bands, enabling better access to their Weyl physics. Thin films MnBi$_2$Te$_4$ provide a rich plethora of topologically distinct quasi-two-dimensional (2D) states, that includes Chern insulator[5-8,12,13] and axion insulator[1,12,14,23] states. The application of external magnetic fields to Mn(Bi$_{1-x}$Sb$_x$)$_2$Te$_4$, therefore, generates an interplay between Chern gaps and Landau levels (LLs) quantization, allowing us to study rich quantum Hall physics that has not yet been fully explored.

## Mn(Bi$_{1-x}$Sb$_x$)$_2$Te$_4$ films at magnetic field B= 0

According to density-functional-theory (DFT), the spin-aligned magnetic configuration of Mn(Bi$_{1-x}$Sb$_x$)$_2$Te$_4$ is a simple type-I (or II depending on lattice parameters[1]) Weyl semimetal with Weyl points at $\pm k_w$ along the $\Gamma - Z$ line. As shown in Fig. 1a, the distance between Weyl points ($2k_w$) decreases with Sb fraction $x$. The bulk Hall conductivity normalized per layer can be expressed as $\frac{\sigma_{xy}^{3D}}{d} = \frac{e^2}{h}\frac{k_w d}{\pi}$, where $k_w$ is the position of Weyl point and $\pi/d$ is the size of Brillouin zone along the $\Gamma - Z$ line with $d$ as the septuple layer (1SL≈ 1.4 nm). The corresponding thin films can be viewed as quasi-2D crystals and are expected to have quantized anomalous Hall conductivities with Chern numbers (C) that increase by one when the film thickness increases by $\Delta t \sim \pi/k_w$[3]. For the case of Sb doping $x$~25% as illustrated by the red curves in Figs. 1a and b, the position of the Weyl point ($k_w \approx 3\pi/25d$) shifts closer to the $\Gamma$ point than at $x$= 0 case, and thus corresponds to the expansion of the C= 1 state to larger film thicknesses. Fig. 1b shows the theoretical thin film gaps versus thickness obtained by fitting the bulk DFT bands to a simplified model[3] as detailed in the supplemental material. The gaps close when a topological phase transition occurs between films with different Chern numbers. The small size of such size-quantization gaps places stringent conditions on the sample quality required for observation of the quantum anomalous Hall effect in the absence of magnetic field. Thus, our focus in this article is on the Chern insulator state in the presence of a magnetic field, which can be observed more consistently than the quantum anomalous Hall effect but is intimately related to the zero magnetic field band topology.

We first examine the low-temperature transport properties of the Mn(Bi$_{1-x}$Sb$_x$)$_2$Te$_4$ thin flakes at zero magnetic field. Fig. 1c plots the four-terminal resistivity ($\rho_{xx}^{CNP}$) measured at the CNP as a function of Mn(Bi$_{1-x}$Sb$_x$)$_2$Te$_4$ thickness for different Sb substitution levels ($x$= 0, 0.20, and 0.26). We refer to a sample as surface-like when its conductivity is thermally activated at low temperatures, suggesting the presence of a bulk energy gap. Samples classified as bulk-like have weaker temperature dependence and are presumed to have disorder-induced bulk states at all energies. Representative temperature-dependent profiles for MnBi$_2$Te$_4$ are shown in Fig. S4, where the samples convert from being surface-like to being bulk-like as thickness increases. The surface-like $\rho_{xx}^{CNP}$ behavior persists to the largest thickness range for the Sb concentration $x$= 0.26 at T= 2K while $\rho_{xx}^{CNP}$ is bulk-like for thickness above 21-SLs. For $x$= 0.20, $\rho_{xx}^{CNP}$ decreases abruptly at thicknesses above 12-SLs. This trend persists for MnBi$_2$Te$_4$. This trend is expected since MnBi$_2$Te$_4$ has bulk n-type doping[12,24], while the substitution of Sb on the Bi sites can shift the Fermi level of the bulk band toward p-type doping, with $x$= 0.26 being closest to CNP[20]. Sb doping at higher concentrations $x$> 0.26 leads to excessive p-type doping and prevents access to CNP in thin flakes[21]. Moving the Fermi level by the Sb to Bi ratio can thus maximize the surface-like regime for probing their quantum transport properties.

Fig. 1d shows a representative $\rho_{xx}$ curve as a function of backgate voltage (V$_{bg}$) for an 18-SL Mn(Bi$_{0.74}$Sb$_{0.26}$)$_2$Te$_4$ device measured at a temperature of 2K. The ambipolar gate-dependent $\rho_{xx}$ suggests a bulk gap with an intrinsic surface state at the CNP (V$_{bg}$~ +3V). The surface carrier density can be tuned to either hole or electron transport by controlling the gate voltage. The field-effect mobility $\mu_{FE} = \frac{1}{C_g}\frac{dG_{xx}}{dV_g}$, where $C_g$ is the gate capacitance (≈80 nF/cm$^2$ for a ~30 nm mica dielectric), $G_{xx}$ (= $1/\rho_{xx}$) is the four-terminal conductance, and $V_g$ is the voltage applied through the graphite gate-electrode, is plotted in Fig. 1d. We see that the electron mobility increases with gate voltage (electron density), while the hole mobility responds weakly and remains small in the low gate voltage (hole density) regime. The mobility is more than one order of magnitude higher mobility for electrons compared to hole carriers. By applying dualgate voltages, the top and bottom surface carrier densities can be modulated to achieve electron mobilities as high as 4000 cm$^2$/Vs at a total carrier density of >5×10$^{11}$ cm$^2$.

**C= 1 state in Mn(Bi$_{1-x}$Sb$_x$)$_2$Te$_4$ films**

The magnetic field-dependent transport properties of Mn(Bi$_{1-x}$Sb$_x$)$_2$Te$_4$ films with the Sb concentrations $x=$ 0, 0.20, and 0.26 for a variety of thicknesses were studied. When a perpendicular magnetic field is applied, Mn(Bi$_{1-x}$Sb$_x$)$_2$Te$_4$ undergoes the magnetic transitions[20,22] from the antiferromagnetic (AFM), to the canted, and finally to the aligned spin-moment configurations. To compare the magnetic field dependence of the Mn(Bi$_{1-x}$Sb$_x$)$_2$Te$_4$ film with different Sb substitutions, we plot the color maps of $\rho_{yx}$ as functions of gate voltage and magnetic field for the 8-SL MnBi$_2$Te$_4$, 10-SL Mn(Bi$_{0.8}$Sb$_{0.2}$)$_2$Te$_4$, and 21-SL Mn(Bi$_{0.74}$Sb$_{0.26}$)$_2$Te$_4$, respectively, in Figs. 2a-c. In general, the spin-flop and spin-flip transitions happen at magnetic fields of ~±2-3T and ~±7T, respectively, are observed in all the samples. The magnetic transition fields are identified by the color line marks depicted in Figs. 2a-c, which agrees with the theoretically calculated values[25]. Line profiles of $\rho_{xx}$ and $\rho_{yx}$ curves for the 8-SL MnBi$_2$Te$_4$, 10-SL Mn(Bi$_{0.8}$Sb$_{0.2}$)$_2$Te$_4$, and 21-SL Mn(Bi$_{0.74}$Sb$_{0.26}$)$_2$Te$_4$ are plotted in Figs. 2d-f. The $\rho_{yx}$ increases sharply with magnetic field in the canted antiferromagnetic (CAFM) phase, and saturates at ~h/e$^2$ as the thin film is driven into the FM phase by magnetic field. The suppression of $\rho_{xx}$ in the FM phase further confirms the development of C= 1 state in all three samples. The interpretation is supported by the gate-dependent $\rho_{xx}$ and $\rho_{yx}$ curves measured at magnetic field of 9T for the three respective samples in Figs. S5a, S6, and extended data Fig. E1, respectively, where the $\rho_{yx}$ plateau and $\rho_{xx}$ minimum can be seen. Despite not being fully quantized, the 21-SL Mn(Bi$_{0.74}$Sb$_{0.26}$)$_2$Te$_4$ film exhibits all the features of the C= 1 state. The anomalous Hall loop at low magnetic field for the 8-SL MnBi$_2$Te$_4$ and 10-SL Mn(Bi$_{0.8}$Sb$_{0.2}$)$_2$Te$_4$ could be due to the uncompensated surface magnetization[15] or antiferromagnetic domain walls[26], whereas no zero field hysteretic behavior observed in the 21-SL Mn(Bi$_{0.74}$Sb$_{0.26}$)$_2$Te$_4$ film. The observed C= 1 states in all three Mn(Bi$_{1-x}$Sb$_x$)$_2$Te$_4$ films with the different Sb substitutions confirm that their spin-alignment configuration is in the topological phase with a Chern insulator gap.

To further evaluate the thickness-dependence of the C= 1 state, we plot $\rho_{xx}$ and $\rho_{yx}$ values as a function of thickness, with the gate voltage tuned to the $\rho_{yx}$ maximum for each thickness, at magnetic field of 9T. Our primary observation is that, despite their similar field-dependent quantization behavior, the C= 1 state prolongs to the larger thickness range with the Sb substitutions as shown in Figs. 2g-i. This trend is consistent with our DFT calculations, which indicate the shift of the Weyl point position by Sb doping. In Fig. 2g, the observation of C= 1 quantum Hall states up to 8-SL MnBi$_2$Te$_4$ agrees with our calculations (Fig. 1b) and the literature[5]. The thickness limit for the C= 1 state extends to 11-SLs for the Mn(Bi$_{0.8}$Sb$_{0.2}$)$_2$Te$_4$ as shown in Fig. 2h. Although no C= 1 state was observed for the MnBi$_2$Te$_4$ and Mn(Bi$_{0.8}$Sb$_{0.2}$)$_2$Te$_4$ beyond 8-SL and 11-SL, respectively, our 11-SL MnBi$_2$Te$_4$ (Fig. S5b) and 16-SL Mn(Bi$_{0.8}$Sb$_{0.2}$)$_2$Te$_4$ (Fig. S7) reveal the developing of $\rho_{yx}$ plateaus with higher Chern numbers near the CNPs. In Fig. 2i, we show the substantially wider thickness range of the C= 1 state resolved for the Mn(Bi$_{0.74}$Sb$_{0.26}$)$_2$Te$_4$. Although this can be somehow related to the extension of the surface-like regime in thin films by the optimal Sb substitutions, according to our calculations in Fig. 1b, the Mn(Bi$_{0.74}$Sb$_{0.26}$)$_2$Te$_4$ films at the given film thickness range should lie in the higher Chern number states.

**Dualgate tuning of Chern states**

To further identify the Chern insulator states in our thicker Mn(Bi$_{0.74}$Sb$_{0.26}$)$_2$Te$_4$, we perform a detailed analysis for these devices in a dualgating platform. Figs. 3a-f compare the dualgate maps of longitudinal conductivity ($\sigma_{xx}$) and Hall resistivity ($\rho_{yx}$) for the Mn(Bi$_{0.74}$Sb0$_{0.26}$)$_2$Te$_4$ devices at the thickness of 21, 18, and 14-SLs, respectively, measured at 9T. As shown in Figs. 3a and b, the 21-SL Mn(Bi$_{0.74}$Sb0$_{0.26}$)$_2$Te$_4$ reveal a clear C= 1 plateau in the dualgate maps. This is also indicated by the $\sigma_{xy}$ ($\sigma_{xx}$) versus V$_{bg}$ line profiles as shown in Fig. 3g. In this device, we observe no other Chern states develop near the CNP besides the C= 1 state at the highest accessible magnetic field of 9T. Careful tracking of the C= 1 state in magnetic field reveals its formation at Fermi energy slightly below the CNP as detailed in extended data Fig. 1. Similar behavior was also observed in the 18-SL Mn(Bi$_{0.74}$Sb0$_{0.26}$)$_2$Te$_4$ (extended data Fig. E2).

    The 18-SL Mn(Bi$_{0.74}$Sb0$_{0.26}$)$_2$Te$_4$ device with higher electron mobility (Fig. 1d) shows more quantization features at 9T. With the dualgating structure, we can access these quantum Hall states by tracing the boundaries of these states with dashed lines in the dualgate maps. The color map of $\sigma_{xx}$ versus dualgate voltages depicted in Fig. 3c shows multiple minima corresponding to the different quantized $\rho_{yx}$ plateaus as indexed in the dualgate map in Fig. 3d, showing the different Chern numbers. The $\sigma_{xy}$ ($\sigma_{xx}$) versus V$_{bg}$ line profiles depicted in Fig. 3h reveal the well-developed C= 3 plateau and the developing C= 1 and 5 quantization states. Tracking the Chern state development at the lower magnetic field reveals an additional C= 2 state in the CAFM phase (extended data Fig. E3). The existence of the C= 2 state in the CAFM phase is also captured by

the $\rho_{xx}$ and $\rho_{xy}$ dualgate maps swept at magnetic field of 6T as shown in extended data Fig. E2. Also, we note that the C= 2 state coincides with the $\rho_{xx}^{CNP}$ at the zero magnetic field.

We further analyze the quantization states in the 14-SL Mn(Bi$_{0.74}$Sb0$_{0.26}$)$_2$Te$_4$. In addition to the C= 1 state, the dualgate maps in Figs. 4e and f show a series of oscillatory $\sigma_{xx}$ minima and $\rho_{xy}$ plateaus, respectively, corresponding to the different quantum Hall states develop at 9T when tuning the dualgate voltages. The linecuts of $\sigma_{xx}$ and $\sigma_{xy}$ as a function of backgate voltage at magnetic field of 9T as depicted in Fig. 4i reveal the quantum Hall plateaus with C= 0, 1, 3, 5, etc. The color maps of $\rho_{xx}$ and $\rho_{yx}$ as functions of magnetic field and backgate voltage (Fig. S8) resolve the fan diagram of Landau levels at odd integer fillings, where the $\rho_{xx}$ minima for each filling factor can be traced down to a single gate voltage corresponding to the CNP. Such feature is similar to the surface states' LL fan diagram in non-magnetic topological insulators[27]. Different from the 18-SL and 21-SL Mn(Bi$_{0.74}$Sb0$_{0.26}$)$_2$Te$_4$, the C= 1 state in the 14-SL Mn(Bi$_{0.74}$Sb0$_{0.26}$)$_2$Te$_4$ at high magnetic field coincides with the CNP when traced down to zero magnetic field. The schematic diagrams inserted in Figs. 3g-i illustrate the surface band structures with different Chern states resolved in the respective thicknesses of the Mn(Bi$_{0.74}$Sb0$_{0.26}$)$_2$Te$_4$ films.

**Anomalous Landau levels in magnetic topological insulators**

To interpret the rich Chern insulator states observed in our Mn(Bi$_{0.74}$Sb$_{0.26}$)$_2$Te$_4$ films, we calculate their LLs spectra using a simplified model[3], in which the 2D massive Dirac cones are coupled by tunneling within and between the compound's septuple-layer building blocks. For explanation purpose, we first present in Fig. 4a the LLs spectra for a non-magnetic TI thin film[28] which has no exchange coupling. The degenerate n= 0 LLs (red curves) lie at the Fermi level due to the non-trivial Berry's phase[29] in addition to the typical surface band LLs (black curves). When the exchange coupling is introduced, the quasi-2D films have discrete finite-length-chain hopping eigenstates, which have no spin-orbit coupling but are spin-split by exchange interactions with the aligned Mn spins. A series of spin-polarized anomalous LLs (red and blue lines) with magnetic field independent energies emerge from the $\vec{k}_\perp = 0$ states as shown in Figs. 4c and d calculated for an 18-SL Mn(Bi$_{0.75}$Sb$_{0.25}$)$_2$Te$_4$ film with intralayer exchange coupling J$_s$ of 30 and 34 meV, respectively. The details of the LLs derivation can be referred to supplementary materials. The Chern numbers of the anomalous LLs are determined by $(N_{E<E_F} - N_{E>E_F})/2$ where $N_{E<E_F}$ ($N_{E>E_F}$) is the total number of subbands below (above) Fermi level. The Chern numbers at B= 0 can be identified from the LLs plots in Figs. 4c and d by determining the filling factor at Fermi level in the limit of zero magnetic field. The Chern number of C= 2 coincides well with the Hall conductivity calculations of the Chern number at B= 0 (Fig. 1b). We see in Figs. 4c and d that the Mn(Bi$_{0.75}$Sb$_{0.25}$)$_2$Te$_4$ films have an interval of carrier-density in which only anomalous LLs are present close to the Fermi level as indicated by the yellow arrows, and the gaps between these levels are large enough to support quantum Hall effects that are robust against disorder. The surface band LLs with n≠ 0 indices (black curves) in conduction-valence pairs move further away from the Fermi level as magnetic fields strengthen and do not contribute to the B= 0 Chern numbers.

A general picture of the relationship between the Chern numbers and exchange coupling is illustrated in Fig. 4b where the surface and bulk spin-splitting bands are plotted. In the case of $J_S = 0$, the number of subbands above and below the Fermi level equals, and thus the Chern number is 0. When the exchange field is turned on ($J_S \neq 0$), the anomalous LLs are spin-polarized with two nearly degenerate surface anomalous LLs labeled with green and red curves in Fig. 4b, the filling factor is thus 1 since the two nearly degenerate surface anomalous LLs are either above or below the Fermi level. When the crossings between up (yellow curves) and down (blue curves) spins happen, the filling factors change by one. The quantization is therefore expected to be observable over a range of filling factors magnitudes centered on 1+ the number of $\vec{k}_\perp = 0$ crossings that occur between up and down spins. When the magnetic field is reversed, the spin of the anomalous LLs is reversed, but the sign of the exchange coupling is also reversed, so the sign of the filling factor range at which strong quantum Hall effects will not change. The sign depends only on the sign of the exchange coupling between the local moments and the Dirac electrons. Our observations, therefore, provide evidence that the anomalous LLs are spin-polarized.

We also note that since the position of Weyl points depends on the exchange splitting[3], which may be decreased by antisite defects leading to a further decrease of $k_w$, and thus may change the Chern numbers and the gaps in the absence of magnetic field[30]. Fig. 4c and 4d examine two cases, one is close to a transition (refer to Figs. S2 and S3 for the dependence of gaps versus various exchange splitting) between two different Chern numbers and hence has a small gap at B= 0 (with $J_s$= 34 meV), and the other is a relatively large Chern gap at B= 0 (with $J_s$= 30 meV). Although the size of the gap at B= 0 is sensitive to details that may vary from sample to sample, the strong magnetic field behavior in which well-spaced anomalous LLs dominate over a finite region of filling factor does not vary too much.

Finally, we compare our experimental results for the Mn(Bi$_{0.74}$Sb$_{0.26}$)$_2$Te$_4$ films with the calculated LLs structure at the similar thickness and Sb doping level. We plot in the extended data Fig. 3 the calculated LL gaps at different filling factors as a function of magnetic field for the Mn(Bi$_{0.75}$Sb$_{0.25}$)$_2$Te$_4$ films. The LL gap size of each filling factor is determined either by gaps between anomalous LLs or by gaps between anomalous and surface band (non-anomalous) LLs. Under a strong magnetic field, the calculated filling factor of C= 1 exhibits the largest LL gap in all three thicknesses of Mn(Bi$_{0.75}$Sb$_{0.25}$)$_2$Te$_4$ film. This explains the experimental observation where the C= 1 state was observed over the wide thickness range in our Mn(Bi$_{0.74}$Sb$_{0.26}$)$_2$Te$_4$ film. One noticeable feature in the case of larger film thickness is that the C= 1 filling shifts below Fermi level and the surface gap at the Fermi level is at filling factor of higher Chern number (refer to Fig. S3 for details). This is consistent with the observation in our 18-SL Mn(Bi$_{0.74}$Sb$_{0.26}$)$_2$Te$_4$ film, indicating that the 18-SL is in a higher Chern number state of C= 2. Moreover, the calculated gap size for the C= 3 state exceeds the C= 2 gap at high magnetic field, which explains the well-developed C= 3 state at 9T at this film thickness. This inevitably supports the development of the anomalous LLs near the Fermi level under strong magnetic field (red shades in Fig. 3g-i). Our results further imply that the Hall quantization states at larger film thickness provide an ideal platform for the realization of the spin-polarized anomalous LLs in thin film Weyl semimetals.

## Summary

In summary, we studied the magnetoelectrical transport of the intrinsic MTI Mn(Bi$_{1-x}$Sb$_x$)$_2$Te$_4$ for $x$= 0, 0.20, and 0.26 by probing Chern quantization states and their relationship with the flake thickness and Sb concentrations. We identified the thickness ranges for surface-like insulating and bulk-like metallic transport regime. The thickness-dependent Hall conductivities, particularly for the C= 1 Chern insulator state in the Mn(Bi$_{1-x}$Sb$_x$)$_2$Te$_4$, show a correlation with the separation of Weyl points as described by our theoretical models, indicating that Mn(Bi$_{1-x}$Sb$_x$)$_2$Te$_4$ behaves as thin film Weyl semimetals. Our transport results for different Sb concentrations highlight the importance of B= 0 Weyl point separation for the Chern state's quantization. Moreover, we showed that the interplay of Chern insulator states at B= 0 and Landau quantization at high magnetic fields gives rise to the intriguing anomalous LLs in Mn(Bi$_{1-x}$Sb$_x$)$_2$Te$_4$. Our work illustrates the complexity of the intertwined topological surface states and ferromagnetism in Landau quantization and thus can serve as a guide to bridge the gap between the 2D Chern insulators and 3D Weyl semimetals.

**Methods**

**Materials.** Mn(Bi$_{1-x}$Sb$_x$)$_2$Te$_4$ bulk crystals at different Sb doping levels were grown by a self-flux growth method[20,24]. Variable thicknesses of Mn(Bi$_{1-x}$Sb$_x$)$_2$Te$_4$ thin flakes were exfoliated from the parent bulk crystals and then transferred into the heterostructures of graphite/hexagonal boron nitride sandwiched layers using a micromanipulator transfer stage. The graphite/hexagonal boron nitride layers serve as the gate-electrode/dielectric layers. The exfoliation and transfer processes were performed in an argon gas-filled glovebox with O$_2$ and H$_2$O levels <1 ppm and <0.1 ppm, respectively, to prevent oxidation in thin flakes. We fabricated the Mn(Bi$_{1-x}$Sb$_x$)$_2$Te$_4$ devices into the Hall bar configuration using a standard electron beam lithography process and metal deposition with Cr/Au (20 nm/60 nm) as the contact electrodes using a CHA Solution electron beam evaporator. The Mn(Bi$_{1-x}$Sb$_x$)$_2$Te$_4$ flakes were protected by polymethyl methacrylate (PMMA) while transporting for lithography and metal deposition processes.

**Measurements.** Low-temperature magnetotransport measurements were performed in a Quantum Design Physical Properties Measurement System (PPMS) in helium-4 circulation (2K-300K) and magnetic field up to 9 T. Two synchronized Stanford Research SR830 lock-in amplifiers at a frequency of 5-8 Hz were used to measure the longitudinal and Hall resistances concurrently on the Mn(Bi$_{1-x}$Sb$_x$)$_2$Te$_4$ devices. The devices were typically sourced with a small AC excitation current of 20-100 nA. Two Keithley 2400 source meters were utilized to source DC gate voltages separately to the top and bottom gate electrodes. Magnetotransport measurements at high magnetic field were carried out in a helium-3 variable temperature insert at a base temperature of 0.4 Kelvin and magnetic field up to 18 tesla based at the National High Magnetic Field laboratory.

**Theoretical calculations.** DFT calculations were performed using Vienna Ab initio Simulation Package (VASP)[31] in which Generalized gradient approximations (GGA) of Perdew-Burke-Ernzerhof (PBE)[32] have been adopted for exchange-correlation potential. On-site correlation on the Mn-3d states is treated by performing DFT + U calculations[33] with U-J as 5.34 eV. The global break condition for the electronic SC-loop is set to be $10^{-7}$ eV and the cutoff energy for the plane wave basis set is 600 eV during the self-consistent (SC) calculations. A 9×9×6 Gamma-centered k-point integration grid was employed with Gaussian broadening factors as 50 meV. In the calculations of bulk Mn(Bi$_{1-x}$Sb$_x$)$_2$Te$_4$, supercells of 2×2×1 unit cell were used to model the doping density of Sb atoms with densities of 0, 25%, 50%, 75%, and 100%. The calculations of Landau levels are based on the coupled Dirac cone model illustrated in the supplemental material, with the parameters estimated from the DFT calculations.

**Data Availability**

The data that support the findings of this study are available from the corresponding authors upon reasonable request.

**Acknowledgements.**

This work was supported by the National Science Foundation the Quantum Leap Big Idea under Grant No. 1936383 and the U.S. Army Research Office MURI program under Grants No. W911NF-20-2-0166 and No. W911NF-16-1-0472. Support for crystal growth and characterization was provided by the National Science Foundation through the Penn State 2D Crystal Consortium-Materials Innovation Platform (2DCC-MIP) under NSF cooperative agreement DMR-2039351. A portion of this work was performed at the National High Magnetic

Field Laboratory, which is supported by the National Science Foundation Cooperative Agreement No. DMR-1644779 and the State of Florida.

**Author Contributions**

S.K.C. and K.L.W. planned the experimental project. S.H.L. and Z.M. prepared the bulk crystals. S.K.C. fabricated the devices and conducted the transport measurements. J.J. helped with transport measurements conducted at National High Magnetic Field laboratory. C.L. and A.H.M. performed the theoretical calculations. S.K.C., L.C., A.H.M. and K.L.W. wrote the manuscript. All authors discussed the results and commented on the manuscript.

**References**

1   Zhang, D. *et al.* Topological Axion States in the Magnetic Insulator ${\mathrm{MnBi}}_{2}{\mathrm{Te}}_{4}$ with the Quantized Magnetoelectric Effect. *Physical Review Letters* **122**, 206401, doi:10.1103/PhysRevLett.122.206401 (2019).

2   Li, J. *et al.* Intrinsic magnetic topological insulators in van der Waals layered MnBi2Te4-family materials. *Science Advances* **5**, eaaw5685, doi:10.1126/sciadv.aaw5685.

3   Lei, C., Chen, S. & MacDonald Allan, H. Magnetized topological insulator multilayers. *Proceedings of the National Academy of Sciences* **117**, 27224-27230, doi:10.1073/pnas.2014004117 (2020).

4   Chowdhury, S., Garrity, K. F. & Tavazza, F. Prediction of Weyl semimetal and antiferromagnetic topological insulator phases in Bi2MnSe4. *npj Computational Materials* **5**, 33, doi:10.1038/s41524-019-0168-1 (2019).

5   Ge, J. *et al.* High-Chern-number and high-temperature quantum Hall effect without Landau levels. *National Science Review* **7**, 1280-1287, doi:10.1093/nsr/nwaa089 (2020).

6   Cai, J. *et al.* Electric control of a canted-antiferromagnetic Chern insulator. *Nature Communications* **13**, 1668, doi:10.1038/s41467-022-29259-8 (2022).

7   Deng, Y. *et al.* Quantum anomalous Hall effect in intrinsic magnetic topological insulator MnBi2Te4. *Science* **367**, 895-900, doi:10.1126/science.aax8156 (2020).

8   Liu, C. *et al.* Magnetic-field-induced robust zero Hall plateau state in MnBi2Te4 Chern insulator. *Nature Communications* **12**, 4647, doi:10.1038/s41467-021-25002-x (2021).

9   Yu, R. *et al.* Quantized Anomalous Hall Effect in Magnetic Topological Insulators. *Science* **329**, 61-64, doi:10.1126/science.1187485 (2010).

10  Chang, C.-Z. *et al.* Experimental Observation of the Quantum Anomalous Hall Effect in a Magnetic Topological Insulator. *Science* **340**, 167-170, doi:10.1126/science.1234414 (2013).

11  Tokura, Y., Yasuda, K. & Tsukazaki, A. Magnetic topological insulators. *Nature Reviews Physics* **1**, 126-143, doi:10.1038/s42254-018-0011-5 (2019).

12  Liu, C. *et al.* Robust axion insulator and Chern insulator phases in a two-dimensional antiferromagnetic topological insulator. *Nature Materials* **19**, 522-527, doi:10.1038/s41563-019-0573-3 (2020).

13  Ovchinnikov, D. *et al.* Intertwined Topological and Magnetic Orders in Atomically Thin Chern Insulator MnBi2Te4. *Nano Letters* **21**, 2544-2550, doi:10.1021/acs.nanolett.0c05117 (2021).

14  Gao, A. *et al.* Layer Hall effect in a 2D topological axion antiferromagnet. *Nature* **595**, 521-525, doi:10.1038/s41586-021-03679-w (2021).

15  Yang, S. *et al.* Odd-Even Layer-Number Effect and Layer-Dependent Magnetic Phase Diagrams in $\mathrm{MnBi}_{2}\mathrm{Te}_{4}$. *Physical Review X* **11**, 011003, doi:10.1103/PhysRevX.11.011003 (2021).

16  Klimovskikh, I. I. *et al.* Tunable 3D/2D magnetism in the (MnBi2Te4)(Bi2Te3)m topological insulators family. *npj Quantum Materials* **5**, 54, doi:10.1038/s41535-020-00255-9 (2020).

17  Rienks, E. D. L. *et al.* Large magnetic gap at the Dirac point in Bi2Te3/MnBi2Te4 heterostructures. *Nature* **576**, 423-428, doi:10.1038/s41586-019-1826-7 (2019).

18  Otrokov, M. M. *et al.* Prediction and observation of an antiferromagnetic topological insulator. *Nature* **576**, 416-422, doi:10.1038/s41586-019-1840-9 (2019).

19  Chen, Y. J. *et al.* Topological Electronic Structure and Its Temperature Evolution in Antiferromagnetic Topological Insulator $\mathrm{MnBi}_{2}\mathrm{Te}_{4}$. *Physical Review X* **9**, 041040, doi:10.1103/PhysRevX.9.041040 (2019).

20  Lee, S. H. *et al.* Evidence for a Magnetic-Field-Induced Ideal Type-II Weyl State in Antiferromagnetic Topological Insulator $\mathrm{Mn}(\mathrm{Bi}_{1-x}\mathrm{Sb}_{x})_{2}\mathrm{Te}_{4}$. *Physical Review X* **11**, 031032, doi:10.1103/PhysRevX.11.031032 (2021).

21  Chen, B. *et al.* Intrinsic magnetic topological insulator phases in the Sb doped MnBi2Te4 bulks and thin flakes. *Nature Communications* **10**, 4469, doi:10.1038/s41467-019-12485-y (2019).

22  Jiang, Q. *et al.* Quantum oscillations in the field-induced ferromagnetic state of $\mathrm{Mn}\mathrm{Bi}_{2-x}\mathrm{Sb}_{x}\mathrm{Te}_{4}$. *Physical Review B* **103**, 205111, doi:10.1103/PhysRevB.103.205111 (2021).

23  Otrokov, M. M. *et al.* Unique Thickness-Dependent Properties of the van der Waals Interlayer Antiferromagnet ${\mathrm{MnBi}}_{2}{\mathrm{Te}}_{4}$ Films. *Physical Review Letters* **122**, 107202, doi:10.1103/PhysRevLett.122.107202 (2019).

24  Lee, S. H. *et al.* Spin scattering and noncollinear spin structure-induced intrinsic anomalous Hall effect in antiferromagnetic topological insulator $\mathrm{MnB}{\mathrm{i}}_{2}\mathrm{T}{\mathrm{e}}_{4}$. *Physical Review Research* **1**, 012011, doi:10.1103/PhysRevResearch.1.012011 (2019).

25  Lei, C., Heinonen, O., MacDonald, A. H. & McQueeney, R. J. Metamagnetism of few-layer topological antiferromagnets. *Physical Review Materials* **5**, 064201, doi:10.1103/PhysRevMaterials.5.064201 (2021).

26  Sass, P. M. *et al.* Magnetic Imaging of Domain Walls in the Antiferromagnetic Topological Insulator MnBi2Te4. *Nano Letters* **20**, 2609-2614, doi:10.1021/acs.nanolett.0c00114 (2020).

27  Chong, S. K., Tsuchikawa, R., Harmer, J., Sparks, T. D. & Deshpande, V. V. Landau Levels of Topologically-Protected Surface States Probed by Dual-Gated Quantum Capacitance. *ACS Nano* **14**, 1158-1165, doi:10.1021/acsnano.9b09192 (2020).

28  Zyuzin, A. A. & Burkov, A. A. Thin topological insulator film in a perpendicular magnetic field. *Physical Review B* **83**, 195413, doi:10.1103/PhysRevB.83.195413 (2011).

29  Pertsova, A., Canali, C. M. & MacDonald, A. H. Thin films of a three-dimensional topological insulator in a strong magnetic field: Microscopic study. *Physical Review B* **91**, 075430, doi:10.1103/PhysRevB.91.075430 (2015).

30  Garnica, M. *et al.* Native point defects and their implications for the Dirac point gap at MnBi2Te4(0001). *npj Quantum Materials* **7**, 7, doi:10.1038/s41535-021-00414-6 (2022).

31  Kresse, G. & Hafner, J. Ab initio molecular dynamics for liquid metals. *Physical Review B* **47**, 558-561, doi:10.1103/PhysRevB.47.558 (1993).

32  Perdew, J. P., Burke, K. & Ernzerhof, M. Generalized Gradient Approximation Made Simple. *Physical Review Letters* **77**, 3865-3868, doi:10.1103/PhysRevLett.77.3865 (1996).

33  Himmetoglu, B., Floris, A., de Gironcoli, S. & Cococcioni, M. Hubbard-corrected DFT energy functionals: The LDA+U description of correlated systems. *International Journal of Quantum Chemistry* **114**, 14-49, doi:https://doi.org/10.1002/qua.24521 (2014).

**Figures**

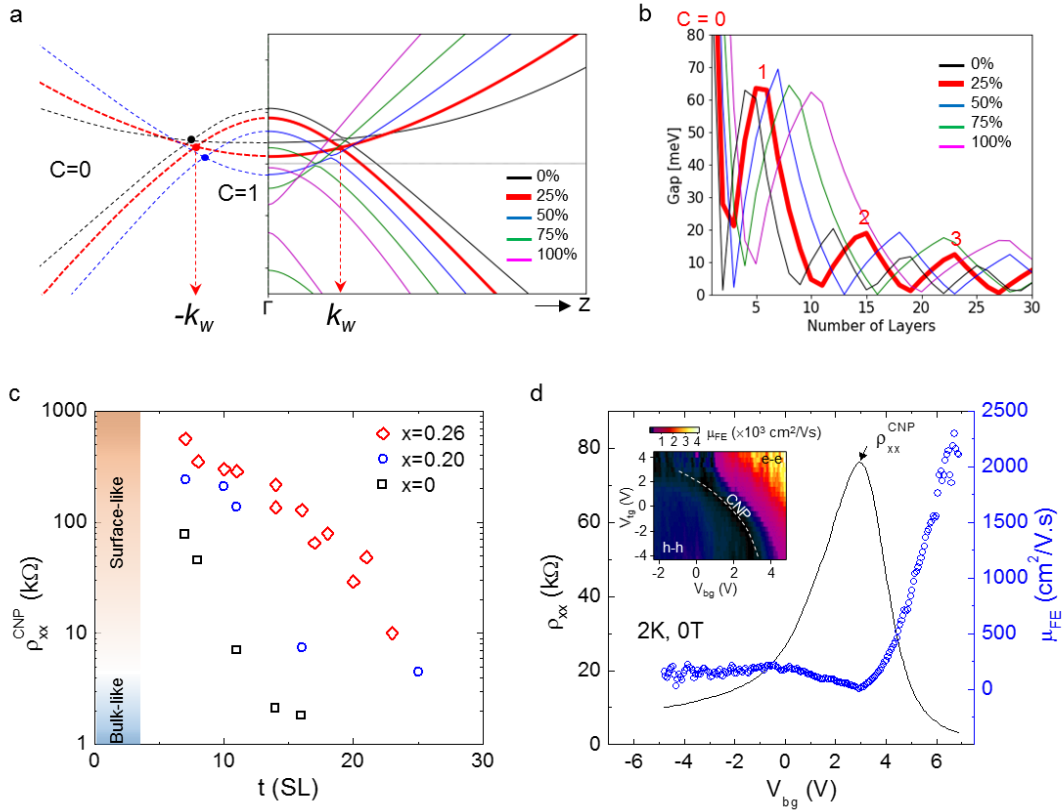

**Figure 1. Zero magnetic field Chern states and transport properties.** (a) DFT calculated band structures for spin-aligned Mn(Bi$_{1-x}$Sb$_x$)$_2$Te$_4$ intrinsic magnetic topological insulators at different Sb substitution levels. The positions of the Weyl point ($k_w$) for $x$= 0%, 25%, and 50% are shown as black, red, and blue nodes, respectively, in the figure. (b) Calculated magnetic exchange gap size as a function of film thickness for the Mn(Bi$_{1-x}$Sb$_x$)$_2$Te$_4$ at different Sb substitution levels. The thickness ranges for Chern numbers, C=0, 1, 2, 3 at $x$= 25% are labeled in the figure. (c) Resistivity at charge neutrality point ($\rho_{xx}^{CNP}$) plot as a function of flake thickness for Mn(Bi$_{1-x}$Sb$_x$)$_2$Te$_4$ at Sb concentrations of $x$= 0, 0.20 and 0.26. $\rho_{xx}^{CNP}$ is defined as the gate-dependent resistivity peak measured at a temperature, T= 2K, at zero magnetic field. (d) $\rho_{xx}$ plot as a function of backgate voltage for a 18-SL Mn(Bi$_{0.74}$Sb$_{0.26}$)$_2$Te$_4$ device measured at temperature, T= 2K, at zero magnetic field. The extracted field effect mobility ($\mu_{FE}$) as a function of backgate voltage is plotted in the secondary y-axis on the right. The inset in (d) is the $\mu_{FE}$ mapping as a function of dualgate voltages taken at 2K.

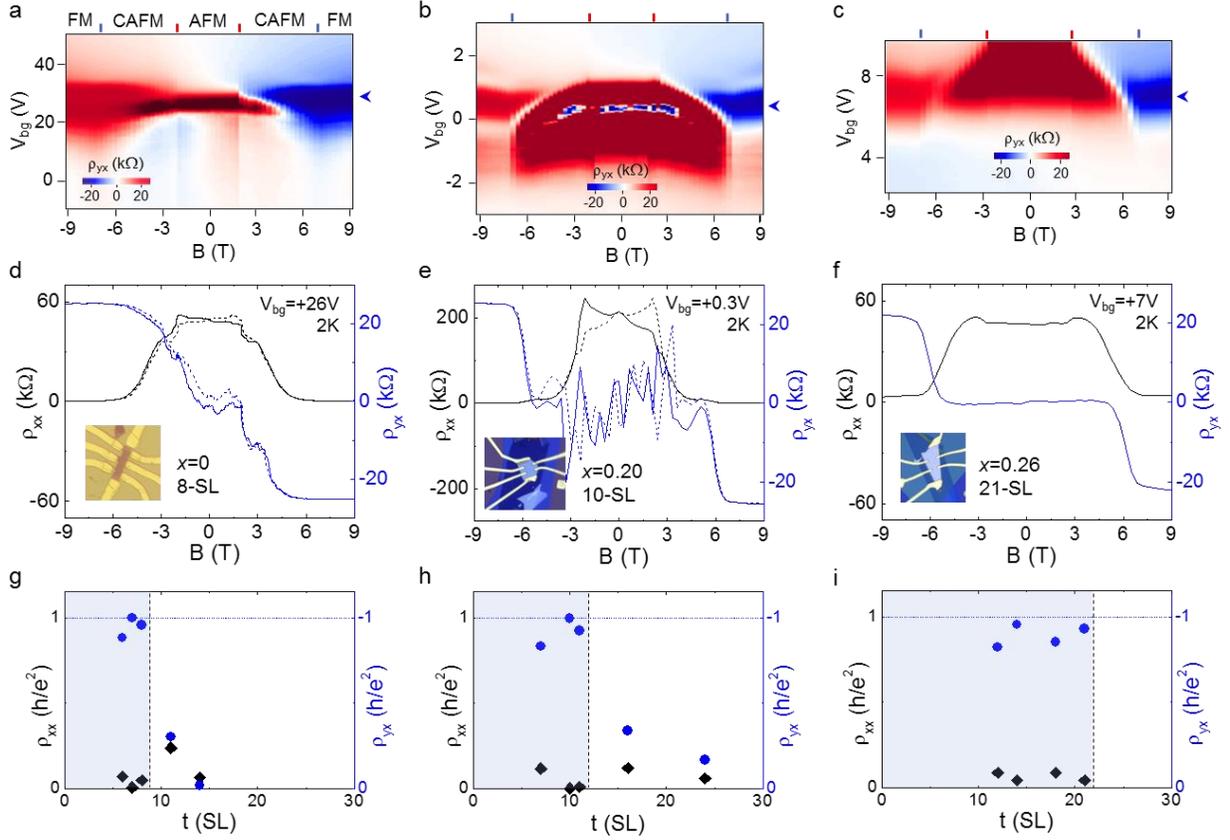

**Figure 2. Magnetic field induced quantization at different Sb substitution levels.** Color maps of $\rho_{yx}$ as a function of magnetic field and gate voltage for (a) $MnBi_2Te_4$, (b) $Mn(Bi_{0.8}Sb_{0.2})_2Te_4$, and (c) $Mn(Bi_{0.74}Sb_{0.26})_2Te_4$ devices at flake thickness of 8-SLs, 10-SLs, and 21-SLs, respectively, measured at temperature of 2K. The red and blue lines mark the transition fields from AFM to CAFM, and CAFM to FM phases, respectively. The $\rho_{xx}$ and $\rho_{yx}$ line profiles in (d), (e), and (f) as a function of magnetic field are extracted from the color maps for the respective devices in (a), (b), and (c) at the gate voltages indicated by the blue arrows. The respective device images are inserted in (d), (e), and (f). The color maps are raw data, whereas the $\rho_{xx}$ and $\rho_{yx}$ line profiles are symmetrized and antisymmetrized, respectively, with respect to the magnetic field. The extracted $\rho_{xx}$ (black rhombus) and $\rho_{yx}$ (blue circle) values at the maximum $\rho_{yx}$ as a function of flake thickness for (g) $MnBi_2Te_4$, (h) $Mn(Bi_{0.8}Sb_{0.2})_2Te_4$, and (i) $Mn(Bi_{0.74}Sb_{0.26})_2Te_4$ devices measured at magnetic field of 9T. The blue color shades in (g)-(i) denote the thickness range where the C= 1 state is observed in the spin-aligned state.

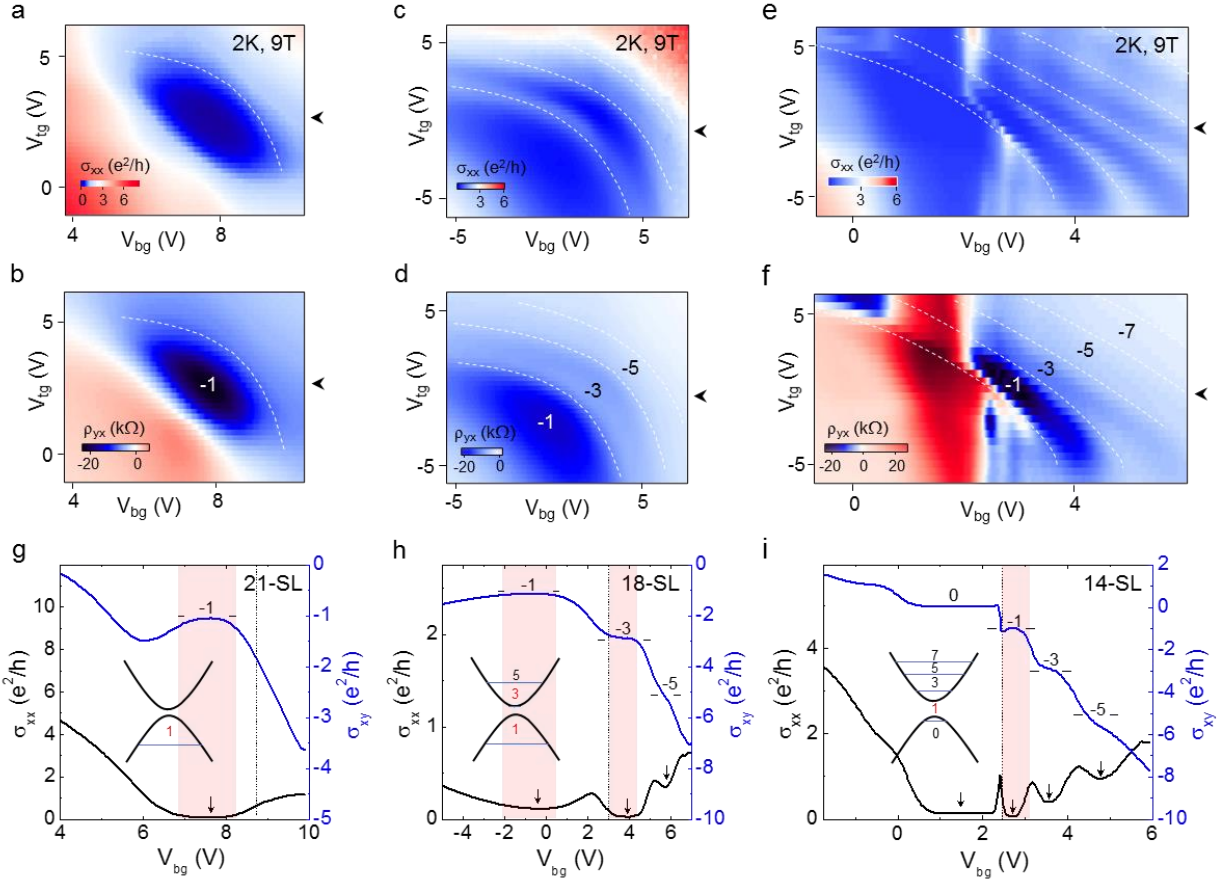

**Figure 3. Tunable Chern states by dualgating.** Color maps of $\sigma_{xx}$ and $\rho_{yx}$ as a function of dualgate voltages for Mn(Bi$_{0.74}$Sb$_{0.26}$)$_2$Te$_4$ at flake thicknesses of (a, b) 21-SLs, (c, d) 18-SLs, and (e, f) 14-SLs, respectively, measured at temperature of 2K and magnetic field of 9T. The white dashed lines in the color maps trace the boundaries of the quantization plateaus with the respective quantum states indexed in the $\rho_{yx}$ maps. Line profiles of $\sigma_{xx}$ and $\sigma_{xy}$ versus backgate voltage curves swept across the charge neutrality as indicated by the black arrows in the color maps for the (g) 21-SLs, (h) 18-SLs, and (i) 14-SLs Mn(Bi$_{0.74}$Sb$_{0.26}$)$_2$Te$_4$. The $\sigma_{xy}$ plateaus and the corresponding $\sigma_{xx}$ minima are indexed to the Chern numbers in (g)-(i). Vertical dashed lines in (g)-(i) mark the backgate voltages corresponding to the $\rho_{xx}^{CNP}$ at zero magnetic field. The surface band structures in (g)-(i) are sketched to illustrate the LL spectra observed for the respective thicknesses at 9T. The red shades in (g)-(i) denote the Chern states forming near the CNP.

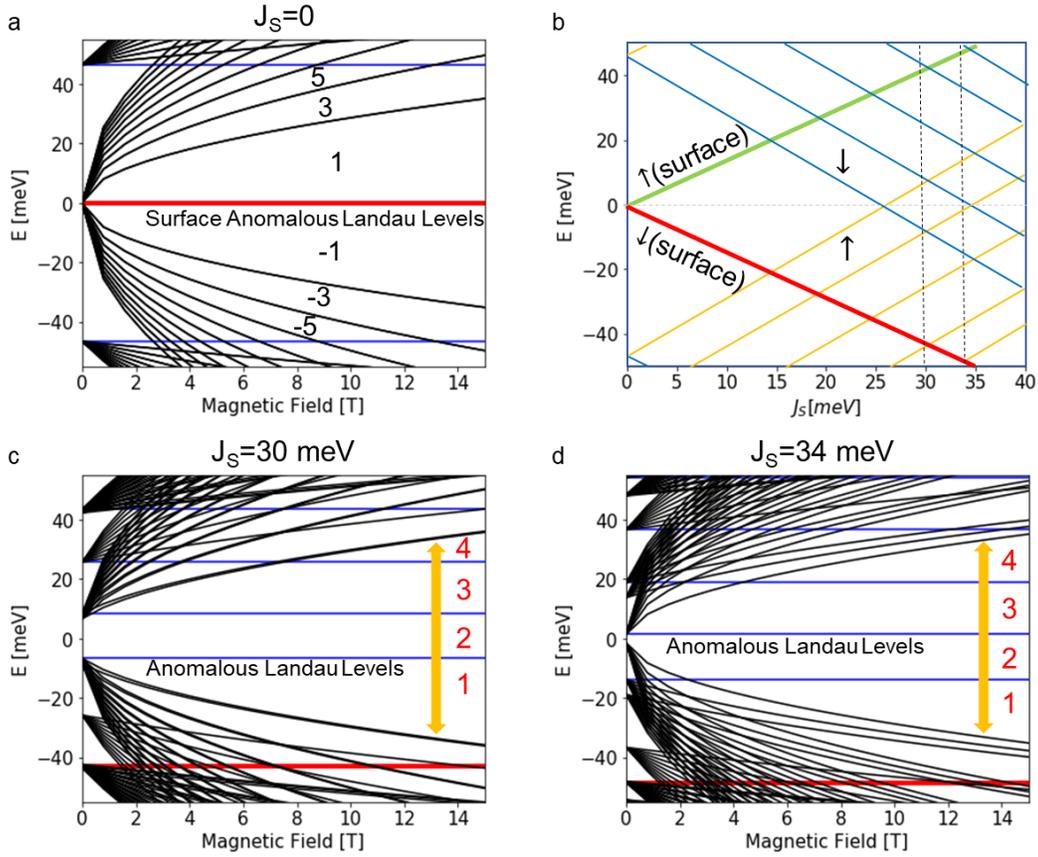

**Figure 4. Spectra of anomalous and non-anomalous Landau levels.** Landau level fan diagrams and filling factors of an 18-SL Mn(Bi$_{0.75}$Sb$_{0.25}$)$_2$Te$_4$ film. (a) The Landau level structures for the case of a non-magnetic thin film (J$_s$= 0). The n≠ 0 non-anomalous Landau levels are plotted with black curves, while the n= 0 anomalous Landau levels whose energies are independent of magnetic field are plotted with blue and red curves. The red curves distinguish anomalous Landau levels that are localized at the surface. (b) Band energies at 2D wavevector $k = 0$ at magnetic field B= 0 versus the same-layer exchange splitting, J$_s$. Spin up (down) states are labeled with orange (blue) color, and the bold green (red) curve is for the spin up (down) state localized at the thin film surfaces. (c) and (d) Landau level structures for thin film with an aligned moment spin configuration at J$_s$= 30 and 34 meV, respectively, labeled in (b) with black vertical dashed lines. Strong quantum Hall states occur when only anomalous Landau levels are close the Fermi level.

**Extended data figures**

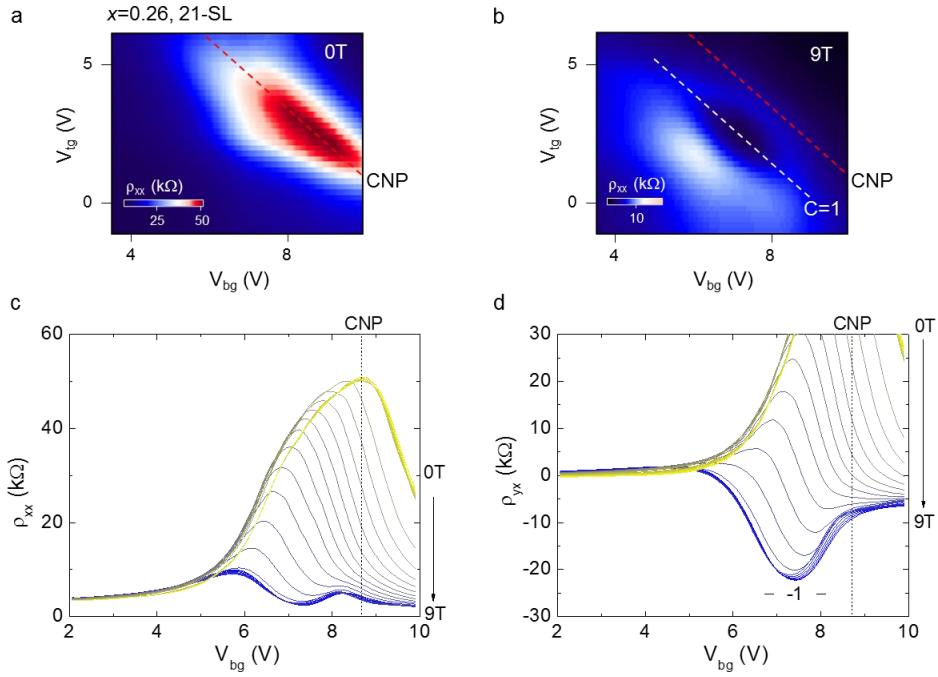

**Extended Data Figure E1. Chern insulator state forming below Fermi level.** Color maps of $\rho_{xx}$ as functions of dualgate voltages measured at magnetic field of (a) 0T and (b) 9T for the 21-SL Mn(Bi$_{0.74}$Sb$_{0.26}$)$_2$Te$_4$ at temperature of 2K. The red and white dashed lines track the charge neutrality of $\rho_{xx}$ maximum at B= 0T and the C= 1 state of $\rho_{xx}$ minimum at B= 9T, respectively. Plots of (c) $\rho_{xx}$ and (d) $\rho_{yx}$ versus backgate voltage measured at different magnetic fields for the 21-SL Mn(Bi$_{0.74}$Sb$_{0.26}$)$_2$Te$_4$. The vertical dashed lines in (c) and (d) indicate the backgate voltage corresponding to the $\rho_{xx}^{CNP}$ at zero magnetic field.

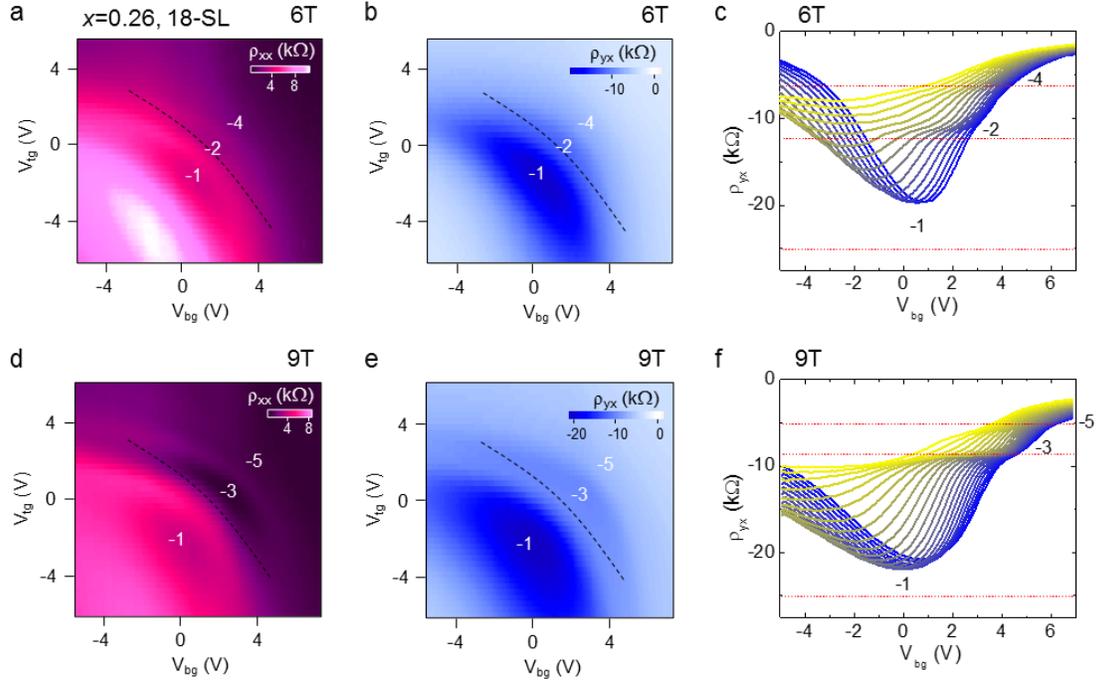

**Extended Data Figure E2. Anomalous Landau levels forming near the charge neutrality.** Color maps of $\rho_{xx}$ and $\rho_{yx}$ as function of dualgate voltages for the 18-SL Mn(Bi$_{0.74}$Sb$_{0.26}$)$_2$Te$_4$ measured at magnetic field of (a, b) 6T, and (d, e) 9T at temperature of 2K. The black dashed lines in the color maps trace the CNP in dualgating as determined from the $\rho_{xx}^{CNP}$ at zero magnetic field. Line profiles of $\rho_{yx}$ as function of backgate voltage taken at different topgate voltages at magnetic field of (c) 6T, and (f) 9T. The developing Chern insulator states of the corresponded $\rho_{xx}$ minima and $\rho_{yx}$ plateaus are indexed in the color maps and line profiles.

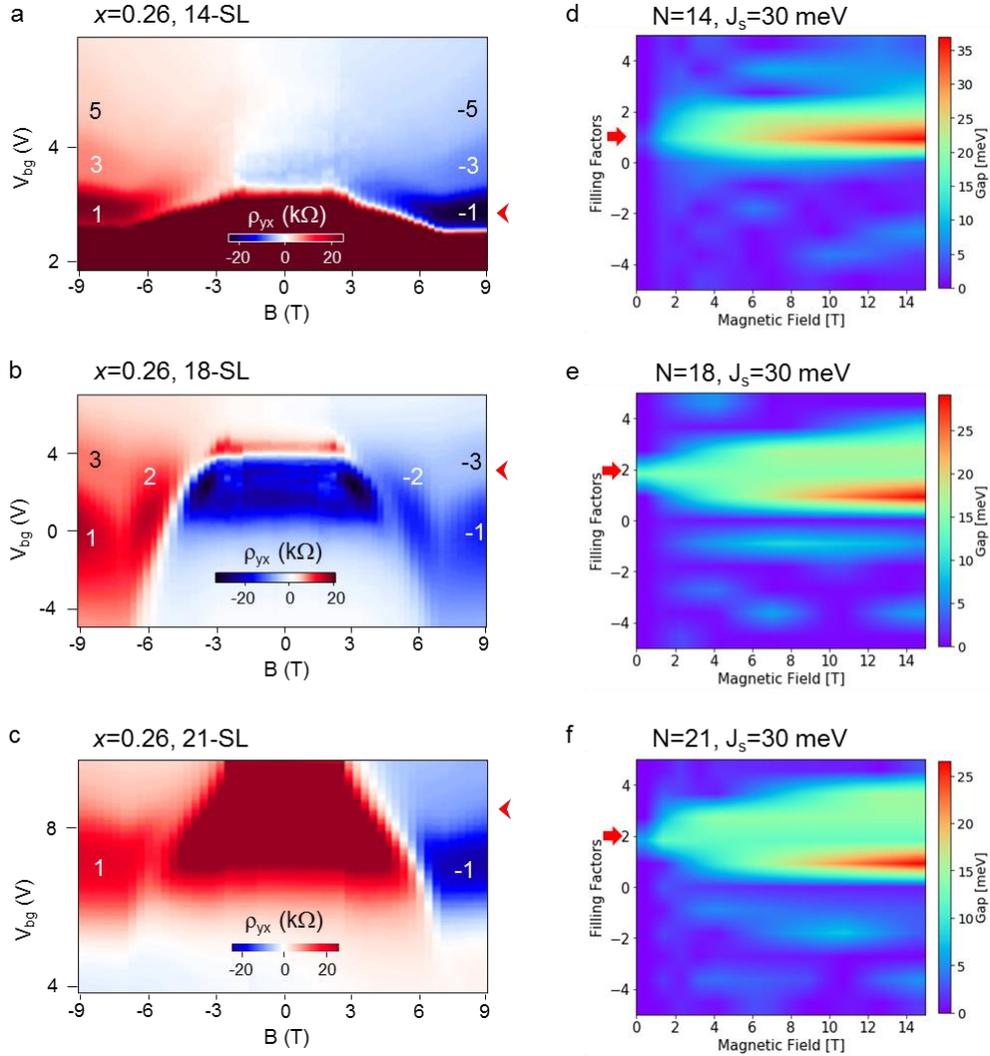

**Extended Data Figure E3. Comparison between the measured and calculated LLs at different thicknesses.** Color maps of $\rho_{yx}$ as functions of magnetic field and backgate voltage for the (a) 14-SL, (b) 18-SL, and (c) 21-SL Mn(Bi$_{0.74}$Sb0$_{0.26}$)$_2$Te$_4$. Contour maps of the Chern gaps as a function of magnetic field for different filling factors extracted from the anomalous and non-anomalous LLs spectra for (d) N= 14-SL, (e) N= 18-SL, and N= 21-SL Mn(Bi$_{1-x}$Sb0$_x$)$_2$Te$_4$ with $x$= 0.25. The C=1 state exhibits the largest gap at strong magnetic field for the different thicknesses. The calculations identify that the Chern number crossing point lies between 14-SLs and 18-SLs for our Mn(Bi$_{0.74}$Sb0$_{0.26}$)$_2$Te$_4$.